\documentclass{aa}
\usepackage{graphicx}
\begin{document}

   \title{Abundance Constraints and Direct Redshift Measurement of the Diffuse X-ray Emission from a Distant Cluster of Galaxies}

   \author{Yasuhiro Hashimoto\inst{1}
          Xavier Barcons\inst{2}
          Hans B\"ohringer\inst{1}
          Andrew C. Fabian\inst{3}
          G\"unther Hasinger\inst{1}
          Vincenzo Mainieri\inst{1} 
          \and
         Hermann Brunner\inst{1}
          }

    \institute{Max-Planck-Institut f\"ur extraterrestrische Physik,
              Giessenbachstrasse
              D-85748 Garching, Germany
          \and
              Instituto de F\'{\i}sica de Cantabraia (CSIC-UC),
              39005 Santander, Spain
          \and
              Institute of Astronomy, 
              Madingley Road, 
              Cambridge, CB3 0HA, UK 
          }

   \offprints{Y.\ Hashimoto,\\ e-mail: hashimot@mpe.mpg.de}

   \date{Received ; accepted}

   \authorrunning{Hashimoto et al.}
   \titlerunning{Abundance of z $>$ 1.0 cluster}

\abstract{
We report on the {\it XMM-Newton} (XMM) observation of RXJ1053.7+5735,
one of the most distant 
X-ray selected clusters of galaxies, which also shows an unusual 
double-lobed X-ray morphology, 
indicative of a possible equal-mass cluster merger.
The cluster was discovered during 
the {\it ROSAT} deep pointings in the direction of the Lockman Hole. 
All XMM Lockman Hole observations (PV, AO-1 \& AO-2 phases)
with the European Photon
Imaging Camera (EPIC) were combined for the analysis,
totaling exposure times
$\sim$ 648 ks, 738 ks, and 758 ks for pn, MOS1, and MOS2, 
respectively. 
With this `deep' dataset,  we
could detect the Fe K line and obtain a strong constraint on 
cluster metallicity, which is difficult to achieve for clusters at z $>$ 1. 
The best-fit abundance is  0.46  $^{+0.11}_{-0.07}$ times the solar value.
The Fe line emission also allows us to directly estimate the redshift of
diffuse gas, with a value  z = 1.14 $^{+0.01}_{-0.01}$.
This is one of the first clusters whose
X-ray redshift is directly measured prior to the secure knowledge of
cluster redshift by optical/NIR  spectroscopy.
We could also estimate the X-ray redshift separately for each 
of the two lobes in the double-lobed structure, and the result is 
consistent with the
two lobes being part of one cluster system
at the same redshift.
Comparison with other metallicity measurements of nearby and
distant clusters shows that there is little evolution
in the ICM metallicity from z $\sim$ 1 to the present.

}

   \maketitle

\section{Introduction}

The heavy element abundances and their distribution in the
intracluster medium (ICM) of galaxy clusters 
provides us with important information on the evolution of the
cluster galaxies, the clusters themselves, and possibly the universe as a whole.
By tracing the evolution of the global metal content of the ICM,
one can obtain useful information on the epoch where the last
major star formation in cluster galaxies took place and
enriched the diffuse cluster gas.
If clusters are fair samples of the universe,
then studies of the evolution of the ICM metallicity also
constrain the star formation history in the universe.
Supernova explosions, which are the main contributor to metal
enrichment, may also provide a significant source of heating for the ICM.
Therefore,
a knowledge of the metal content of galaxy clusters
allows one to estimate the total amount of energy supplied by
supernova explosions (e.g. Pipino et al. 2002).

X-ray spectroscopic observations provide a powerful means to probe the
metal content of the diffuse gas in clusters.
In the late 70's, the iron line emission was discovered in the X-ray
spectra of galaxy clusters (e.g. Serlemitsos et al. 1977).
It was later found that
the ICM in z $\sim$ 0 clusters of galaxies
has been enriched in iron to approximately $1/3$ of the solar value
(e.g. Edge \& Stewart 1991).
{\it ASCA} and {\it BeppoSAX} showed that the metallicity of the ICM
stays constant at its local value
$ Z = 0.3Z_{\odot} $ out to z $\sim$ 0.4 
(e.g. Mushotzky \& Loewenstein 1997; 
Allen \& Fabian 1998; Della Ceca et al 2000; 
Matsumoto et al 2000; Ettori, Allen \& Fabian 2001).
These observations clearly indicate that a significant amount of the metals
produced by supernovae were injected by z $\geq$ 0.4.

With Chandra and {\it XMM-Newton} (XMM) observations on distant clusters,
the evolution of chemical abundances 
beyond z = 0.4 can be investigated,
thus extending the previous analysis based
on {\it ASCA} and {\it BeppoSAX} data (e.g. Jones et al. 2003; Arnaud et al. 2002;
Jeltema et al. 2001; Maughan et al. 2003; Worrall et al. 2003; 
Holden et al. 2002).
Tozzi et al. (2003) analyzed mostly-Chandra data of 18 distant clusters
with redshift 0.3 $<$ z $<$ 1.3. They report, using a combined spectral fit of
cluster subsamples in different redshift bins, that the mean Fe
abundance at z $\sim$ 0.8 is $ Z = 0.25^{+0.04}_{-0.06} Z_{\odot} $, 
consistent with the local metallicity value. 
Unfortunately, for clusters at redshift z $>$ 1,
they could only 
provide  weak constraints for the metallicity 
($ Z = 0.21^{+0.12}_{-0.05} Z_{\odot}$ )even after combining four clusters.

The cluster RXJ1053.7+5735,
which shows an unusual double-lobed
X-ray morphology, 
was discovered during our
deep (1.31 Msec) {\it ROSAT} HRI pointings
(\cite{Hget98}), 
in the direction of the ``Lockman Hole",
a line of sight with exceptionally low HI column density.
The angular size of the source is 1.7 $\times$ 0.7 arcmin$^2$
and 
two lobes are approximately 1 arcmin apart.
The total X-ray flux of the entire source 
in the 0.5-2.0 keV band 
is 2 $\times$ 10$^{-14}$ erg cm$^{-2}$ s$^{-1}$
(\cite{Hget98b}).
Our subsequent deep optical/NIR imaging follow-ups (V $<$ 26.5, R $<$ 25, I $<$ 25,
K $<$ 20.5) with LRIS and NIRC on Keck,
and the Calar Alto Omega Prime camera
revealed a bright 7 arcsecond-long arc with an
integrated magnitude of R=21.4 as well as
an overdensity of galaxies in both X-ray lobes
(e.g. \cite{Thet01}).
Further Keck LRIS/NIRSPEC spectroscopic observations on the bright arc
and one of the brightest galaxies near the arc showed that the former
is a lensed galaxy at a redshift z = 2.57
while the latter is at a redshift of z = 1.263 (\cite{Thet01}).  
Deep VRIzJHK photometry data 
produced concordant
photometric redshifts for more than 10 objects
at redshift of z $\sim$ 1.1-1.3, confirming
that at least the eastern lobe is a massive cluster at high redshift.
The improbability of chance alignment and similarity of colors for the galaxies in the
two X-ray lobes were consistent with the western lobe also being at z $\sim$ 
1.1-1.3 (\cite{Thet01}).
Meanwhile, the X-ray data of the XMM observation (with a total 
effective exposure time
$\sim$ 100 ks)
performed during the PV phase for this cluster 
were analyzed 
(Hashimoto et al. 2002), yielding   
a best-fit temperature of 4.9 $^{+1.5}_{-0.9}$ keV,
while the metallicity was poorly constrained 
with an upper limit on the iron abundance of 0.62$Z_{\odot}$.

Here we report on a deep $\sim$ 1 Msec XMM observation of
the cluster RXJ1053.7+5735, based on the 
data obtained during the 
PV, AO-1, \& AO-2 phases.
The paper is organized as follows. In Sec. 2, we briefly
describe the observation and data reduction.
In Sec. 3, we first present spectroscopic analysis of the
entire cluster, then of each lobe.
In Sec. 4, we present a discussion of the metallicity evolution, as well as  
the comparison between the X-ray  redshift and optical/infrared redshifts.
All uncertainties quoted are at the 68\% confidence levels for a
one interesting parameter fit. 
Throughout the paper, we use $H_{o}$ = 65 km s$^{-1}$ Mpc$^{-1}$,
$\Omega_{m}$=0.3, and $\Omega_{\Lambda}$=0.7 

\section{Observations and Data Reduction}
The cluster 
RXJ1053.7+5735 was observed 
as a part of the XMM Lockman Hole observation
with 
European Photon
Imaging Camera (EPIC) during the period April 27-May 19, 2000 
(rev. 70, 71, 73, 74, and 81)
for $\sim$ 170 ks (PV phase), 
the period Oct. 25-Nov. 4, 2001 
(rev. 344, 345, and 349) for $\sim$ 199 ks (AO-1),
and the period Oct. 15-Dec. 7, 2002 (rev. 522-528, and 544-548) for $\sim$ 920 ks (AO-2).
Both EPIC cameras, pn and MOS, were operated in Full Frame mode and 
with the medium optical blocking filter for the bulk of time,
except for the PV observation, where the thin filter was used for
pn, and thin and thick filters were  used for MOS.
The light curves were visually inspected, and we
discarded the data corresponding to the
high particle background ($>$ 8 ct/s for the pn, and  $>$ 5.2 ct/s
for MOS1 and MOS2 inside the entire fov in the 0.5-10keV band).
The data with pattern $>$ 4 and pattern $>$ 12
were also excluded for pn and MOS, respectively,
retaining only single and double events for pn, and
single to quadruple events for MOS.
We further removed those events with low spectra quality
(i.e. FLAG $>$0).
The remaining  exposure times after this cleaning
is approximately 648 ks, 738 ks and 758 ks for pn, MOS1, and MOS2, 
respectively.

\section{RESULTS}

\begin{figure}
 \resizebox{\hsize}{!}{\includegraphics[bb=40 212 575 581,clip]{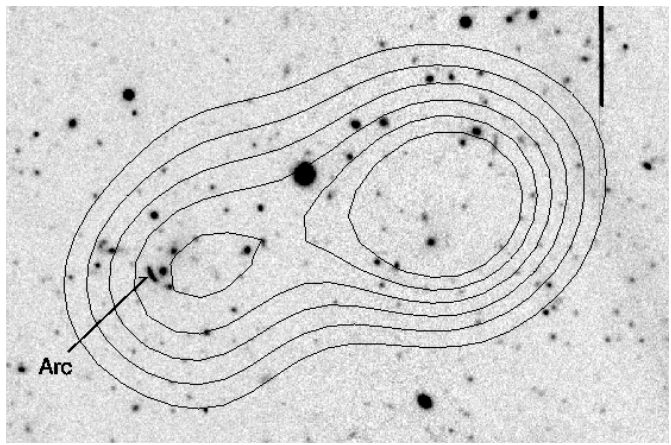}}
 \caption{ 
The contours of XMM image of the cluster RXJ1053.7+5735
overlaid on a  CFHT I band image.
The image was created by combining all events 
in the 0.2-8 keV band
from three
(pn, MOS1, \& MOS2) cameras.
North is up and East is left.
The image is 2\farcm3 $\times$ 1\farcm5 on a side.
The raw data were smoothed with a Gaussian with $\sigma$ = 7\arcsec.
The lowest contour is 1.9 counts arcsec$^{-2}$ and
the contour interval is 0.2 counts arcsec$^{-2}$.
}
\label{FigTemp}
\end{figure}

\subsection{Overall Cluster Analysis}

The spectra were extracted from an elliptical region centered at the
cluster image with 
a semi-major/minor axis 1\farcm2 \& 0\farcm6, respectively
The backgrounds were estimated from a co-centric
annulus region surrounding  the cluster
with 1\farcm3 (inner) and 2\farcm1 (outer) radii,
after removing point sources.
We obtain  4605 and 4305 net source counts (in the 0.2-10.0 keV band)
for pn and MOS(1+2), respectively. 

To correct for vignetting effects, we used the SAS V5.4.1 task EVIGWEIGHT 
(Arnaud et al. 2002),
so that we can use on-axis effective areas.
This task  computes a weight coefficient for each event, defined as the ratio of
the effective area at the photon position and energy to the central 
effective area at that energy, including filter transmission
and quantum efficiency.
Redistribution matrices are generated by the SAS RMFGEN for each pointing,
filter, and detector, and then combined for joint analysis.
We regrouped all the source spectra so that no energy bin had fewer than 50 counts. 
We fitted the spectra with a
redshifted MEKAL model (e.g. Mewe et al. 1985; Kaastra \& Mewe 1993)
 using the 
XSPEC V11.0 (Arnaud 1996).
We fit pn and MOS data (MOS1 and MOS2 are merged) jointly with common temperature, abundance, 
and redshift, while using separate normalization for each dataset, 
and fixing the
column density of neutral 
hydrogen $N_{\rm {H}}$ at a value of 5.6 $\times$ 10$^{19}$ cm$^{-2}$
(Dickey \& Lockman 1990).

\begin{figure}[h]
\resizebox{\hsize}{!}{\includegraphics[angle=-90,bb=0 30 555 700,clip]{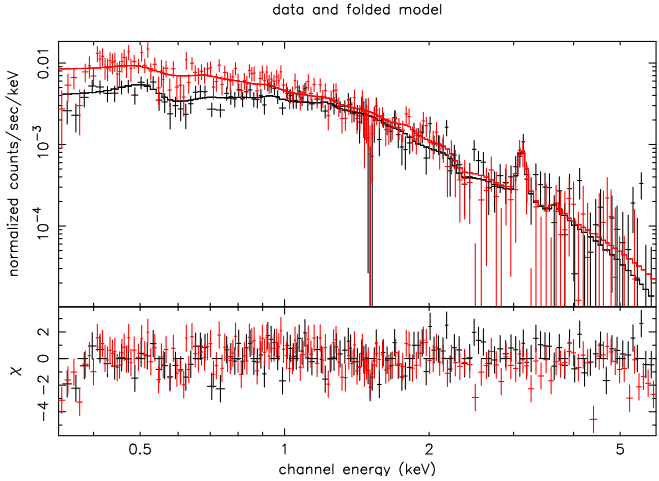}}
\caption{ Rebinned spectra, residuals, and best-fit models 
for cluster RXJ1053.7+5735
 with MOS1+2 (lower spectrum) , and pn (upper spectrum) cameras.
The crosses are the observed spectra and the solid lines denote
 best-fit models.
}
\label{FigTemp}
\end{figure}
\begin{figure}[h]
 \resizebox{\hsize}{!}{\includegraphics[angle=-90,clip]{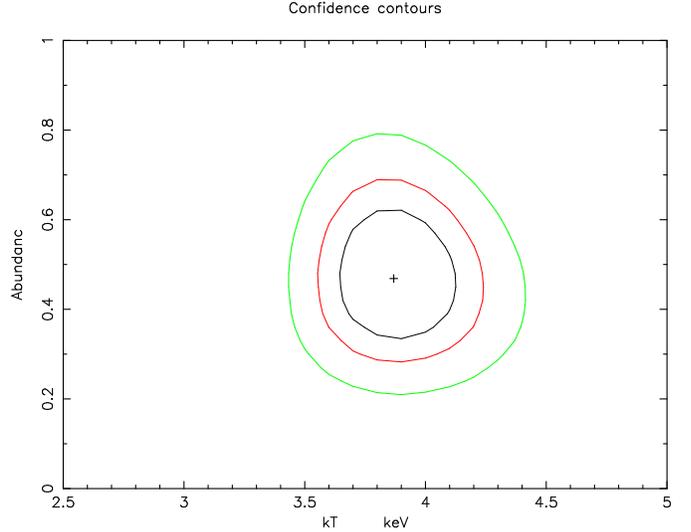}}
 \caption{
Two-dimensional $\chi^{2}$ contours at 68.3\%, 90\%, and 99\%
confidence levels ($\Delta\chi^{2}$=2.30, 4.61 and 9.21) for the
temperature $kT$ and the abundance $Z/Z_{\odot} $ of the 
cluster RXJ1053.7+5735. Both the temperature and the abundance are obtained
from  the joint fit to the pn and MOS data.
}
\label{FigTemp2}
\end{figure}

In Fig. 2, the results of the best-fits are shown, while
Fig. 3 shows the two-parameter $\chi^{2}$ contours for the cluster metallicity
and X-ray temperature.
The best-fit values  based on a simultaneous fit 
are $kT$ =
3.9 $^{+0.2}_{-0.2}$ keV,
an abundance of 0.46 $^{+0.11}_{-0.07}$ times the solar value,
and a redshift z = 1.14 $^{+0.01}_{-0.01}$. 
Using the best-fit model parameters, we derived an
unabsorbed (0.2-10) keV
flux of $f_{0.2-10}$ =
2.8 $\times$
10$^{-14}$ erg cm $^{-2}$ s$^{-1}$,
corresponding to a
luminosity in the cluster rest frame of
$L_{0.2-10}$ =
2.4 $\times$
10$^{44}$ erg s$^{-1}$ and a bolometric luminosity of
$L_{bol}$ =
2.8 $\times$
10$^{44}$ erg s$^{-1}$.
Note that the best-fit redshift is consistent with
the redshift range  derived by the photometric redshift technique
(Thompson et al. 2001),
while it is significantly lower than
the NIR spectroscopic redshift (z=1.26) of one galaxy in the eastern lobe
 near the gravitational arc. We will further comment about  this point
in \S 4.

\subsection{Substructure}

As illustrated in Fig. 1, the cluster RXJ1053.7+5735 
shows a clearly double-lobed structure,
which extends roughly in the east-west direction.
Using a circular (radius = 0\farcm47) extraction region centered at each lobe,
X-ray characteristics are investigated for the substructure.

Fig. 4 shows X-ray spectra and residuals for the eastern lobe (top panel)
and for the western lobe (bottom panel).
Fig. 5 shows two-dimensional $\chi^{2}$ contours at 68.3\%, 90\%, and 99\%
confidence levels ($\Delta\chi^{2}$=2.30, 4.61 and 9.21) for the
eastern and western lobes of the cluster temperature and redshift, 
obtained from the joint fit to the pn and MOS data.
Our fitting for each lobe shows that $kT$ is
3.4 $^{+0.2}_{-0.1}$ keV \& 4.4 $^{+0.3}_{-0.3}$ keV, 
the redshift is 1.16 $^{+0.02}_{-0.03}$ \& 1.14 $^{+0.02}_{-0.01}$,
and the abundance is 0.41 $^{+0.18}_{-0.16}$ \& 0.62 $^{+0.15}_{-0.14}$
for eastern and western lobe, respectively 
(Table 1).
Note that the redshifts of two lobes agree within a statistical limit,
consistent with the interpretation that two lobes are at close 
physical distance. 

\begin{figure}[h]
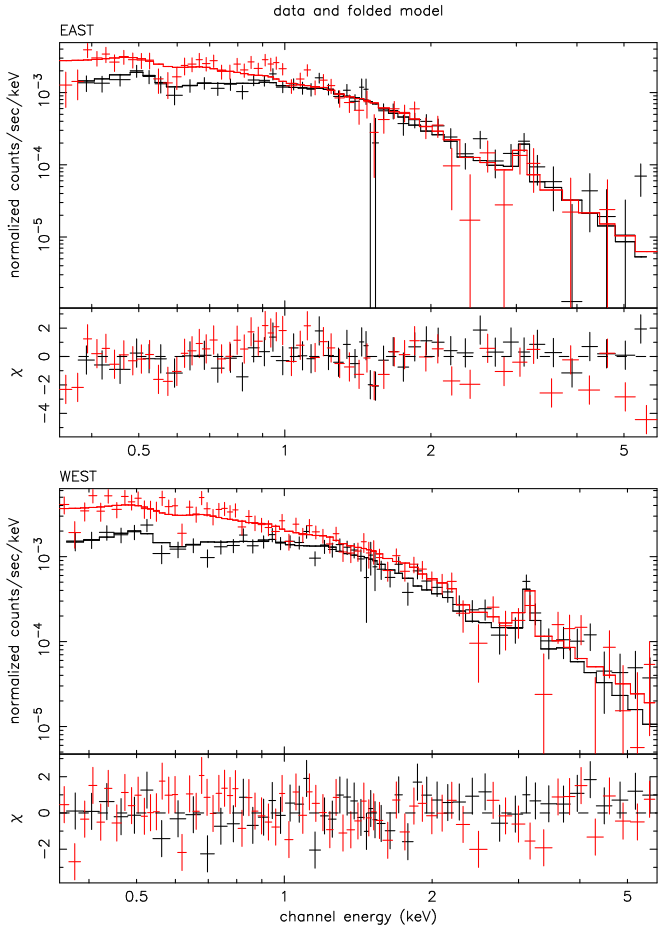

\resizebox{\hsize}{!}{\includegraphics[angle=-90,bb=0 30 539 700,clip]{lh_mos.070-548.screened.eastlobe.grp+lh_pn.070-548.screened.eastlobe.grp.mekal.annulusbg.lda.ps}}
\resizebox{\hsize}{!}{\includegraphics[angle=-90,bb=90 30 570 700,clip]{lh_mos.070-548.screened.westlobe.grp+lh_pn.070-548.screened.westlobe.grp.mekal.annulusbg.lda.ps}}
\caption{
X-ray spectrum and residuals for the east lobe (top panel), and for the
west lobe (bottom panel).
}
\label{FigTemp}

\end{figure}

\begin{table}[h]
\caption[]{Comparison between the two lobes (pn+MOS1+MOS2)}
\begin{tabular}{lrrrl}
\hline
\hline
\noalign{\smallskip}
Lobe         & $kT$  & $Z/Z_{\odot} $   & Redshift &  R.$\chi^{2}$(dof) \\
              & (keV) &               & (z) &    \\ 
\noalign{\smallskip}
\hline
\noalign{\smallskip}
East  & 3.4$^{+0.2}_{-0.1}$ &  0.41$^{+0.18}_{-0.16}$ & 1.16$^{+0.02}_{-0.03}$ & 1.47(99) \\
West  & 4.4$^{+0.3}_{-0.3}$ &  0.62$^{+0.15}_{-0.14}$ & 1.14$^{+0.02}_{-0.01}$ & 1.00(119) \\
ALL   & 3.9$^{+0.2}_{-0.2}$ &  0.46$^{+0.11}_{-0.07}$ & 1.14$^{+0.01}_{-0.01}$ & 1.17(301) \\
\noalign{\smallskip}
\hline
\end{tabular}
\end{table}

\begin{figure}[h]
\resizebox{\hsize}{!}{\includegraphics[angle=-90,clip]{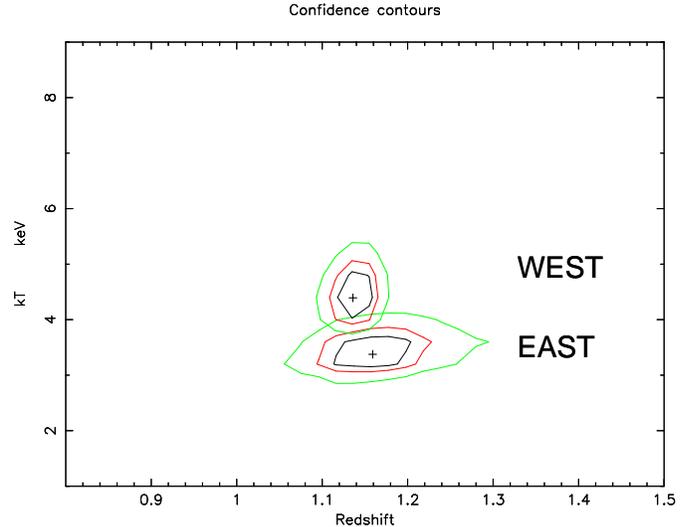}}
\caption{
Two-dimensional $\chi^{2}$ contours at 68.3\%, 90\%, and 99\%
confidence levels ($\Delta\chi^{2}$=2.30, 4.61 and 9.21) for the
temperature $kT$ and the redshift of the eastern and western lobes
of cluster RXJ1053.7+5735.
}
\label{FigTemp}
\end{figure}

\begin{figure}
 \resizebox{\hsize}{!}{\includegraphics[angle=90]{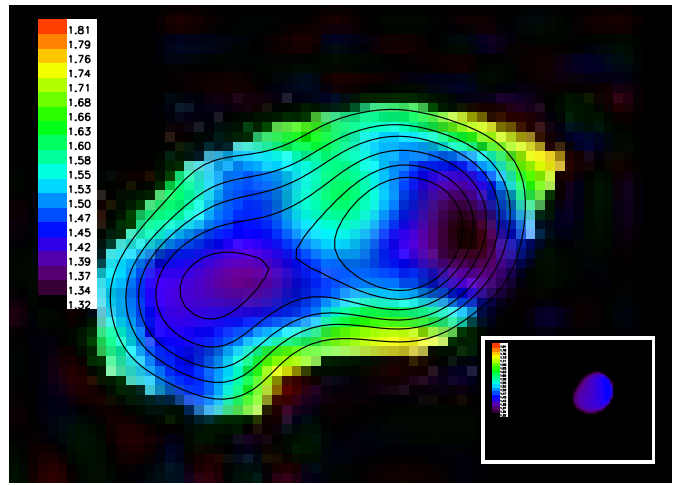}}
 \caption{
The X-ray hardness ratio map where
the hard band image (1-8 keV) is divided by soft band image
(0.2-1 keV).
Both images are separately smoothed,
before the division, with a Gaussian with $\sigma$ = 7\arcsec.
The full-band (0.2-8 keV)
contours are overlaid on the hardness ratio map.
Illustratively, the ratio = 1.8
here roughly corresponds to $kT \sim$ 9-10 keV, while
the value 1.3 corresponds to  $kT \sim$ 3-4 keV.
Inset shows the hardness ratio map of the $\beta$ model 
convolved 
with the detector point spread function (PSF) 
at the position of the cluster
(westlobe), demonstrating 
that the energy dependence of the detector PSF
has negligible effect on the cluster hardness ratio map.
}
\label{FigTemp}
\end{figure}

To investigate the indication of detailed 
temperature variations in the cluster, 
we created an X-ray `hardness ratio' map, whose structure may be 
interpreted as temperature variations.
The X-ray hardness ratio map is created, first, by making
hard (1-8 keV) and soft (0.2-1 keV) band images, then by 
dividing the hard band image by the soft band image.
Two images are separately  smoothed
before the division with a Gaussian with $\sigma$ = 7\arcsec. 
This  map is shown in Fig. 6.
The full-band (0.2-8 keV) 
contours are overlaid on the hardness map.
Inset shows the hardness ratio map of the $\beta$ model 
($\beta$=0.67, $r_{c}$=17\farcs2)
convolved with the detector PSF
at the position of the cluster
(westlobe). The inset covers the 
region of the sky identical to the main figure of Fig. 6.
The profile of each pointing and each detector
in 0.2-1, 1-2, 2-5, and  5-8 keV bands are separately
created by the SAS V.6.0 emldetect task with a parameter extentmodel=beta.
The hard band (1-8 kev) is split into three bands
and later combined, in order 
to reflect the cluster X-ray flux in each band.
The scale bar in the inset is shown as a reference for the image-size
(the color-scales are the same as the main figure).
The inset demonstrates that the energy dependence of the detector PSF
has negligible effect on the cluster hardness ratio map, where the bulk of
the cluster emission is below 5 keV.
Fig. 6 shows that the coldest part of the western lobe may be slightly 
off-centered toward west, while the temperature structures of the eastern lobe
is aligned with respect to the full-band contours.
The region between the two lobes
seems to have similar temperature with the outer region of each
lobe, and seems hotter than the bulk of the  cluster temperature,
however, the influence from the cluster-cluster  collision is
unclear based on our data. We will further discuss about this point
in \S 4.

\section{Discussion}

\subsection{Cluster Redshift}
The redshift of the cluster diffuse gas
directly estimated by fitting the X-ray data is
z = 1.14 $^{+0.01}_{-0.01}$.
This value is within  the cluster redshift-range obtained
with the photometric redshift technique. However, it is significantly
lower than
a redshift (z=1.26), of one cD-like galaxy near the eastern-lobe arc,
obtained by near infrared spectroscopy.
The infrared redshift is practically defined by
one line detected at 1.485 $\mu$m, which Thompson et al. (2001) interpreted
as H$\alpha$ in light of its photometric redshift.

If the redshift of this cD-like galaxy were at z = 1.14, as
indicated by the X-ray  redshift, then this emission
line detected in the cD-like galaxy
would not correspond to any prominent line.
Moreover, no line is detected
at 1.404 $\mu$m where possible H$\alpha$ line should be located,
if the galaxy is indeed at z = 1.14.
Therefore, it is unlikely that redshift of this particular galaxy is  at
z=1.14.

The simplest explanation for this redshift  discrepancy is:
the cD-like galaxy is in the background of the cluster,
and therefore, not a member of the cluster
(relative velocity $\sim 17,000\, {\rm km}\, {\rm s}^{-1}$).
Even with the apparent proximity of the cD-like galaxy
to the gravitational arc and
to the center of the X-ray contour, this explanation is indeed possible.
In fact, there is another cD galaxy candidate equally close to the
center of the X-ray contour.  We obtained DEIMOS/Keck spectroscopy
of this galaxy and our preliminary analysis shows that
the redshift of the galaxy is
consistent  with that of  X-ray diffuse gas.

\subsection{Dynamical State of the Cluster}

The fact that the X-ray morphology of RXJ1053.7+5735 is
double-lobed suggests that we may be witnessing a possible
equal-mass merger event at z $>$ 1.
We could  estimate the X-ray redshift separately for each lobe,
and the result is consistent with the
interpretation that two lobes are a part of one  cluster system
at the same redshift, although the dynamical state of the system
cannot be definitively known from the redshift-information alone.

As for the Tx distribution, as shown in Fig. 6, 
the inter-lobe region seems  hotter than the bulk of each lobe, 
however it is not clear from the figure that 
the temperature in that region is actually enhanced, by the mechanism
such as a shock induced by the cluster `collision'.
Relatively abrupt temperature boundary between the low temperature region
in the western lobe
and the inter-lobe region may be similar with
the `cold fronts' observed in local clusters
(e.g. Markevitch et al. 2000; Vikhlinin et al. 2001).
In these `cold fronts', a sharp temperature gradient forms when a
cooler gas blob propagates in a hotter ambient gas.
The cold western lobe spot which shows a steep temperature gradient
relative to an intermediate hot
cluster-intersection region might be interpreted as one such `cold front'.
Note, however, that our abrupt temperature boundary does not seem to match
exactly the densest part of the gas (as shown by the contours)
and therefore this tentative interpretation has to be taken with some caution.

Relatively high Fe abundance of RXJ1053.7+5735 is
somewhat surprising, particularly considering its redshift.
However, it may be explained by the possible existence of `densed core' 
(or `colling flow') in the cluster, and its relation to the Fe abundance
(e.g. Allen \& Fabian 1998; see also De Grandi \& Molendi 2001).
The hardness ratio map of RXJ1053.7+5735 may indicate that
the temperature is cooler at the
center of each lobe, which is consistent with  the cooling flow 
interpretation.
However, this `densed core' feature  has been 
considered to be associated with 
a relaxed dynamical state of clusters
(e.g. McGlynn \& Fabian 1984; Edge, Stewart, \& Fabian 1992).
If the  two lobes of RXJ1053.7+5735 are indeed interacting, 
and are in the process of nearly-equal-mass merger,
it would be unlikely that  this dynamical state exists
inside the cluster, and therefore unlikely to have the `densed core' profile.
Unfortunately, the spatial resolution of our data is not 
sufficient enough to directly investigate the core profile of the cluster
RXJ1053.7+5735, and therefore, the implied dynamical state of the cluster
remains unknown.
Future X-ray observations with high spatial resolution,
as well as the  optical/NIR spectroscopic observations
of individual cluster members, are needed
for more definitive statement about the dynamical state of the cluster.

\subsection{Abundance vs. Redshift}
In Fig. 7, we compare our newly measured metallicity for RX J1053.7+5735 
with other high redshift clusters
observed by XMM or {\it Chandra} from Tozzi et al. (2003). 
For comparison, the low redshift samples from
Mushotzky \& Loewenstein (1997), and Irwin \& Bregman (2001) are also plotted. 
The black square denotes the new abundance for RX J1053.7+5735.
The eastern and western lobes of the cluster are also separately shown
(labeled as ``East Lobe'' and ``West Lobe").
The error bars  on the abundance  are one dimensional 68\% confidence range.
Some 90\% errors quoted in the literature are converted to 68\% error
by multiplying by a factor 1/1.6.
 
\begin{figure}[h]
 \resizebox{\hsize}{!}{\includegraphics[angle=90,bb=30 30 580 700,clip]{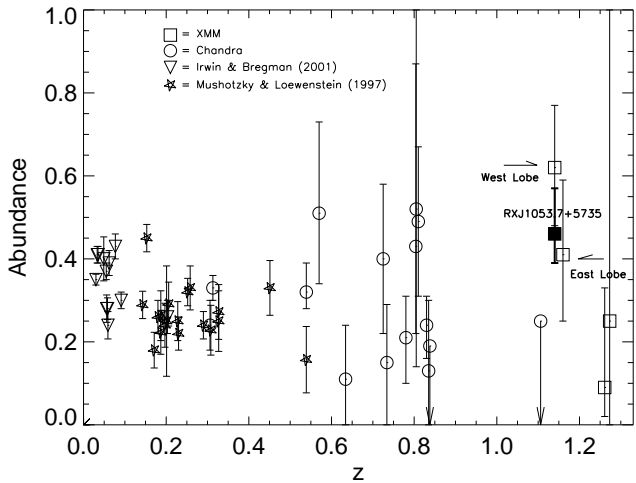}}
\caption{
New XMM abundance  and its 1$\sigma$ error (68.3\%) for RXJ1053.7+5735 plotted
with high redshift clusters from Tozzi et al. (2003).
For comparison, the low redshift samples from
Mushotzky \& Loewenstein (1997), and Irwin \& Bregman (2001) are also plotted.
}
\label{FigTemp}
\end{figure}

Several studies report a hint of
a weak dependence of global Fe abundance on the temperature 
in such a way that lower temperature clusters tend to have larger Fe 
abundances than hot clusters 
(e.g. Yamashita 1992; Fabian et al. 1994; Mushotzky \& Loewenstein 1997;
Matsumoto et al. 2000; 
Tozzi et al. 2003, 
see however Fukazawa et al. 1998; Fukazawa et al. 2000). 
This dependence may be related to the correlation between the
core profile and Fe abundance, which indicates that more `densed core'
or `cooling flow' clusters show the higher Fe abundance
(e.g. Allen \& Fabian 1998).
Regardless of the origin of this Fe dependence on the cluster temperature,
Tozzi et al. (2003) subdivided their cluster sample according to the 
temperature (the `hot' cluster subsample with $kT >$ 5 keV and  
the `cold' cluster subsample with $kT <$ 5 keV). 
They investigated the abundance vs. redshift relation separately for
each subsample using a combined fit across all of the subsample clusters in 
different redshift bins, and   
reported that 
at least the high temperature subsample 
reveals no sign of the evolution in their {\it average} abundances up to $z \sim$ 1.2.

\begin{figure}[h]
 \resizebox{\hsize}{!}{\includegraphics[angle=90,bb=30 30 580 700,clip]{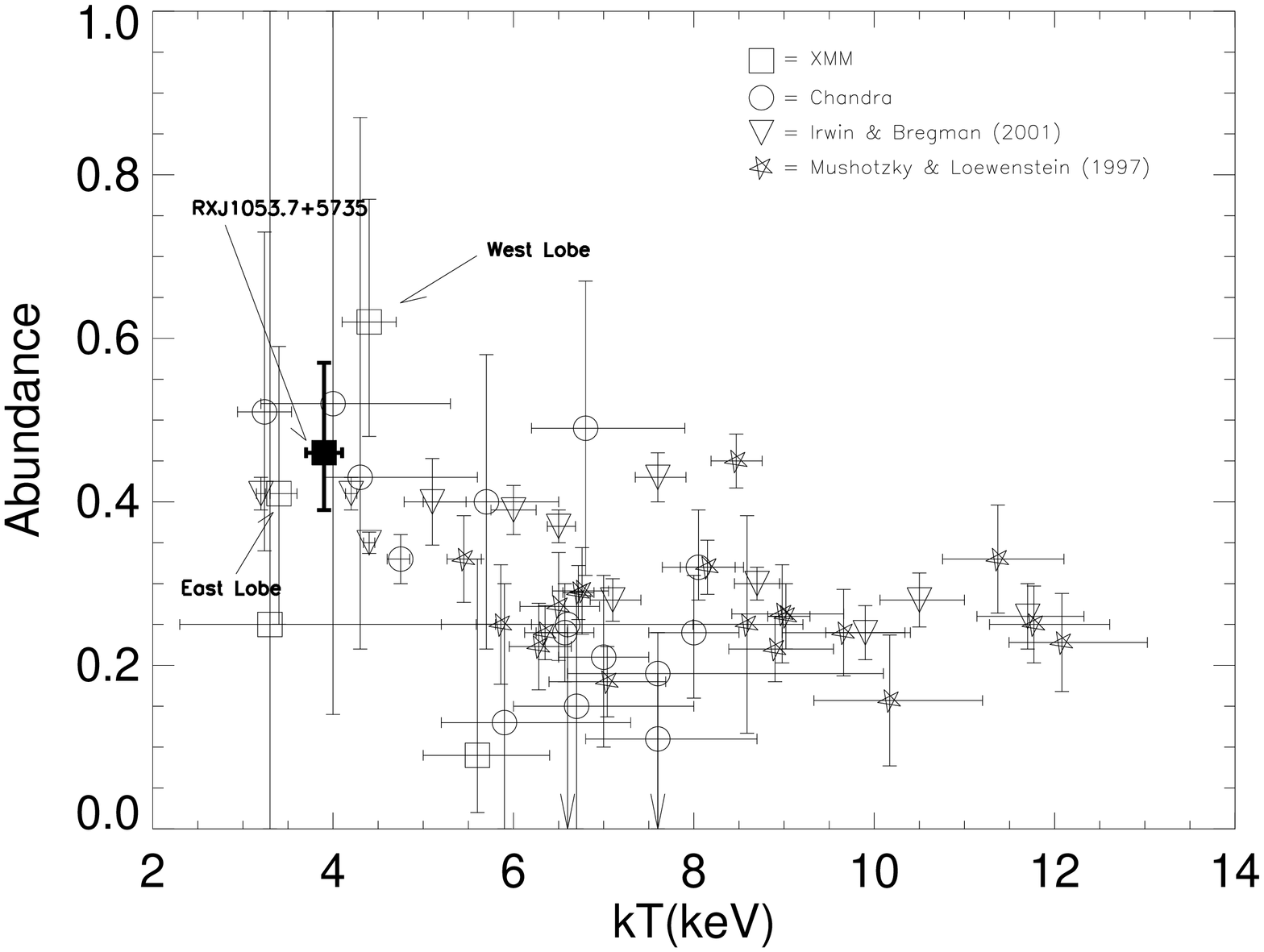}}
 \caption{
New XMM abundance and Tx  and their 1$\sigma$ errors (68.3\%) for RXJ1053.7+5735 plotted
with high redshift clusters from Tozzi et al. (2003).
The low redshift samples from
Mushotzky \& Loewenstein (1997), and Irwin \& Bregman (2001) are also plotted.
}
\label{FigTemp}
\end{figure}

\begin{figure}[h]
 \resizebox{\hsize}{!}{\includegraphics[angle=90,bb=30 30 580 700,clip]{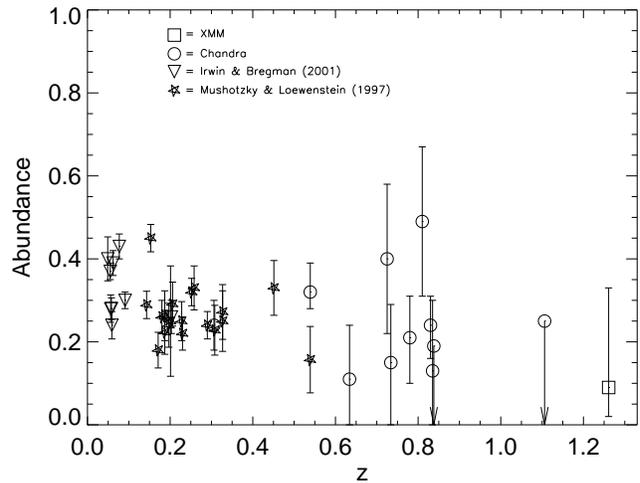}}
 \resizebox{\hsize}{!}{\includegraphics[angle=90,bb=30 30 580 700,clip]{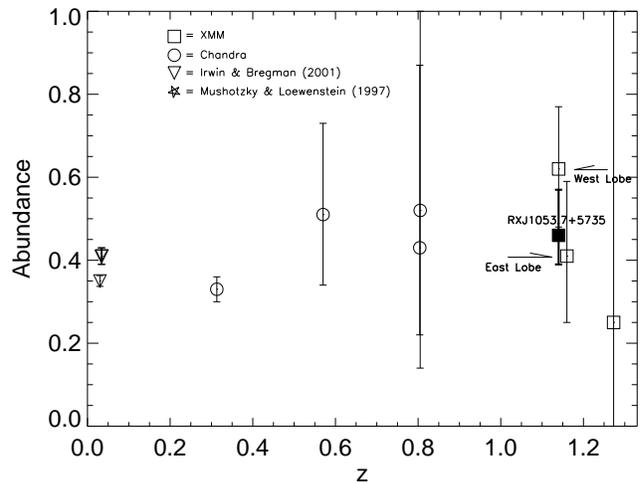}}
 \caption{
Abundance vs Redshift for two subsets of clusters defined by X-ray temperature.
Top panel: Clusters with $kT >$ 5 keV, Bottom panel: clusters with $kT <$ 5 keV
}
\label{FigTemp}
\end{figure}

Fig. 8 shows the abundance vs. temperature relation for the same sample
as Fig. 7. The symbols are identical with Fig. 7.
The error bars on the temperature are again 68\% confidence range.
The cluster RXJ1053.7+5735 roughly  follows a weak trend that
lower temperature clusters tend to have larger iron 
abundances than hot clusters. 
Fig. 9 shows the abundance vs. temperature relation for the `hotter'
clusters ($kT >$ 5 keV: Fig. 9a), and
`cooler' clusters ($kT < $ 5 keV: Fig. 9b).
Again, the black square denotes the new abundance for RX J1053.7+5735,
with eastern and western lobes separately shown.
For both subsamples, there is no statistically-significant redshift
dependence, while the mean abundance is somewhat higher for the cooler
subsample. 
The result is consistent with no-evolution of the Fe abundance  at z
$\sim$ 1.2,
supporting the scenario that much of the enrichment and  
star formation took place earlier.

The cluster RX J1053.7+5735 is now one of the first clusters whose
X-ray redshift is directly measured prior to the secure knowledge of
the cluster redshift obtained with optical/NIR  spectroscopy.
Deep pointings with the current X-ray satellites
provide a powerful means to study the high-redshift clusters,
enabling us to directly measure the redshifts,
and to probe the metal contents, of the diffuse gas in the clusters.
The high-throughput X-ray spectroscopic mission in the future, such as
Con-X and XEUS, should make these studies  almost a routine.

\begin{acknowledgements}
We thank Pat Henry for a careful reading of the paper and useful comments.
XB acknowledges partial financial support by the Spanish Ministerio
de Ciencia y Tecnolog\'{\i}a, under project AYA2000-1690.
We acknowledge referee's comments which improved the manuscript.
\end{acknowledgements}


\begin{thebibliography}{}

\bibitem[Allen \& Fabian 1998]{AlFa98}
Allen S.W., \& Fabian A.C. 1998, MNRAS, 297, L63


\bibitem[Arnaud  1996]{Ar96}
Arnaud K.A., 1996, In ASP Conference Series, Vol 101, 17

\bibitem[Arnaud et al. 2002]{Aret02}
Arnaud M., Majerowicz S., Neumann D.M., et al. 2002, A\&A 390, 27 








\bibitem[De Grandi \& Molendi 2001]{DeMo01}
De Grandi S., \& Molendi S. 2001, ApJ 551, 153

\bibitem[Della Ceca et al. 2000]{Deet00}
Della Ceca R., Scaramella R., Gioia I.M., Rosati P., Fiore F., \& Squires G.
2000 A\&A 353, 498

\bibitem[Dickey and Lockman 1990]{DiLo90}
Dickey J.M., \& Lockman F.J. 1990 ARA\&A 28, 215


\bibitem[Edge \& Stewart 1991]{EdSt91}
Edge A.C., \& Stewart G.C. 1991, MNRAS, 252, 428

\bibitem[Edge, Stewart, \& Fabian 1992]{EdStFa92}
Edge A.C., Stewart G.C., \& Fabian A.C., 1992, MNRAS, 255, 431


\bibitem[Ettori et al. 2001]{Etet01}
Ettori S., Allen S.W., \& Fabian A.C., 2001, MNRAS 322, 187



\bibitem[Fabian et al. 1994]{Faet94}
Fabian A.C., Crawford C.S., Edge A.C., \& Mushotzky R.F., 1994, MNRAS, 267, 779 



\bibitem[Fukazawa et al. 1998]{Fuet98}
Fukazawa Y., Makishima K., Tamura T, et al. 1998 PASJ, 50, 187

\bibitem[Fukazawa et al. 2000]{Fuet00}
Fukazawa Y., Makishima K., Tamura T, et al. 2000 MNRAS, 313, 21 


\bibitem[Hashimoto et al. 2002]{Haet02}
Hashimoto, Y., Hasinger G., Arnaud M., Rosati P., \& Miyaji T., 2002, A\&A 381, 841

\bibitem[Hasinger et al. 1998a]{Hget98}
Hasinger G., Burg R., Giacconi R., et al., 1998a, A\&A 329, 482 

\bibitem[Hasinger et al. 1998b]{Hget98b}
Hasinger G., Giacconi R., Gunn J.E., et al., 1998b, A\&A 340, L27






\bibitem[Holden et al. 2002]{Hoet02}
Holden B.P., Stanford S.A., Squires G.K., et al. 2002, AJ 124, 33


\bibitem[Irwin \& Bregman 2001]{IrBr01}
Irwin J.A., \& Bregman J.N., 2001, ApJ 546, 150

\bibitem[Jeltema et al.]{Jeet01}
Jeltema T.E., Canizares C.R., Bautz M.W., Malm M.R., Donahue M., Garmire G.P.
2001, ApJ 562, 124

\bibitem[Jones et al. 2003]{Joet03}
Jones C., Maughan B.J., Ebeling H. et al. 2003, astro-ph/0304264



\bibitem[Kaastra \& Mewe 1993 ]{KaMe93}
Kaastra J.S., \& Mewe R. 1993, A\&AS, 97, 443



\bibitem[Matsumoto et al. 2000]{Maet00}
Matsumoto, H., Tsuru, T., Fukazawa, Y. et al. 2000, PASJ 52, 153

\bibitem[Maughan et al. 2003]{Maet03}
Maughan B.J., Jones L.R., Ebeling H. et al. 2003, ApJ 587, 589


\bibitem[Mewe et al. 1985]{Meet85}
Mewe R., Gronenschild E.H.B.M., \& van den Oord G.H.J 1985, A\&AS, 62, 197



\bibitem[Markevitch et al. 2000]{Maet00}
Markevitch, M. et al. 2000, ApJ 541, 542 

\bibitem[McGlynn \& Fabian 1984]{McFa84}
McGlynn T.A., \& Fabian A.C., 1984, MNRAS, 208, 709




\bibitem[Mushotzky \& Loewenstein 1997]{MuLo97}
Mushotzky R., Loewenstein M., 1997, ApJ 481, L63






\bibitem[Pipino et al. 2002]{Piet02}
Pipino A., Matteucci F., Borgani S., et al., 2002, NewA 7, 227








\bibitem[Serlemitsos et al. 1977]{Seet77}
Serlemitsos P.J, Smith B.W., Boldt E.A., Holt S.S., \& Swank J.H., 1977, ApJ 211, L63





\bibitem[Thompson et al. 2001]{Thet01} 
Thompson D., Pozzetti L., Hasinger G, et al., 2001, A\&A 377, 778

\bibitem[Tozzi et al. 2003]{Toet03} 
Tozzi P., Rosati P., Ettori S., et al. 2003, ApJ 593, 705 



\bibitem[Vikhlinin et al. 2001]{Viet01} 
Vikhlinin A., Markevitch M., Murray S.S., 2001, ApJ 551, 160


\bibitem[Worrall \& Birkinshaw 2003]{WoBi03} 
Worrall D.M., \& Birkinshaw M., 2003, MNRAS 340, 1261



\bibitem[Yamashita 1992]{Ya92} 
Yamashita K., 1992, in Frontiers of X-ray Astronomy, ed. Tanaka Y. \& Koyama K. (Tokyo: Universal Academy Press)



\end{thebibliography}
\end{document}